\documentclass[twocolumn,prl,showpacs,superscriptaddress]{revtex4}
\usepackage{amssymb}
\usepackage{amsmath}
\usepackage{graphicx}
\usepackage{subfigure}
\usepackage{natbib}
\usepackage{epsfig}
\usepackage{amsfonts}
\usepackage{mathrsfs}
\usepackage{ulem}
\usepackage{color}
\usepackage[toc,page,title,titletoc,header]{appendix}

\begin{document}

%%%%%%%%%%%%%%Equation version%%%%%%%%%%

\newcommand {\beq}{\begin{equation}}
\newcommand {\eeq}{\end{equation}}
\newcommand {\beqa}{\begin{eqnarray}}
\newcommand {\eeqa}{\end{eqnarray}}         %Equation version
\newcommand {\beqs}{\begin{eqnarray*}}
\newcommand {\eeqs}{\end{eqnarray*}}
\newcommand {\bds}{\begin{displaymath}}
\newcommand {\eds}{\end{displaymath}}
\newcommand {\n}{\nonumber\\}
\newcommand {\nn}{\nonumber}
%%%%%%%%%%%%%%%

%%%%%%%%%%%%%Reference version%%%%%%%%%%%%%%%%
\newcommand {\bebb}{\begin{thebibliography}}
\newcommand {\eebb}{\end{thebibliography}}      %Reference version
\newcommand {\bbit}{\bibitem}

\def\dz{\frac{d}{dz}}

\newcommand{\cA}{{\cal A}}
\newcommand{\cB}{{\cal B}}
\newcommand{\cC}{{\cal C}}
\newcommand{\cD}{{\cal D}}
\newcommand{\cF}{{\cal F}}
\newcommand{\cG}{{\cal G}}
\newcommand{\cH}{{\cal H}}
\newcommand{\cI}{{\cal I}}
\newcommand{\cK}{{\cal K}}
\newcommand{\cM}{{\cal M}}
\newcommand{\cN}{{\cal N}}
\newcommand{\cP}{{\cal P}}
\newcommand{\cQ}{{\cal Q}}
\newcommand{\cR}{{\cal R}}
\newcommand{\cS}{{\cal S}}
\newcommand{\cT}{{\cal T}}
\newcommand{\cU}{{\cal U}}
\newcommand{\cV}{{\cal V}}
\newcommand{\cW}{{\cal W}}
\newcommand{\cY}{{\cal Y}}
\newcommand{\cZ}{{\cal Z}}
%%%%%%%%%%%%%%%%%%%%%%%%%%%%%%%%%
%%%%%%%%%%%%%%%%%%%%%%%%%%%%%%%%%
\newcommand{\evn}{{\overline{0}}}
\newcommand{\odd}{{\overline{1}}}

%\newcommand
%%%%%%%%%%%%%%%%%%%%%%%%%%%
%  Useful symbols         %
%%%%%%%%%%%%%%%%%%%%%%%%%%%
\def\a{\alpha}
\def\bt{\beta}

\def\d{\delta}
\def\D{\Delta}

\def\eps{\epsilon}
\def\veps{\varepsilon}

\def\G{\Gamma}
\def\gm{\gamma}

\def\k{\kappa}

\def\l{\lambda}
\def\L{\Lambda}

\def\o{\omega}
\def\O{\Omega}

\def\vph{\varphi}

\def\vpi{\varpi}

\def\s{\sigma}
\def\S{\Sigma}
\def\vsgm{\varsigma}

\def\t{\theta}
\def\T{\Theta}
\def\vt{\vartheta}

\def\z{\zeta}

%%%%%%%%%%%%%%%For over letters%%%%%%%%%%%%%%%%%%%%
\def\tl{\tilde}  %%%%%%%%%%For over letters%%%%%%%%%
\def\wht{\widehat}
\def\arr{\arrow}
%%%%%%%%%%%%%%%%%%%%%%%%%%%%%%%%%%%%%%%%%%%
\def\empst{\emptyset}
\def\prm{\prime}
\def\exst{\exists}
\def\el{\ell}
\def\nba{\nabla}
\def\p{\partial}
\def\bksh{\backslash}
\def\sd{\surd}
\def\nft{\infty}
\def\fal{\forall}
\def\cpd{\coprod}
\def\pd{\prod}
\def\bgp{\bigup}
\def\bgc{\bigcap}
\def\prt{\propto}

%%%%%%%%%%%%%%%%%%%%brackets
\def\la{\langle}
\def\ra{\rangle}
\def\rb{\rbrace}
\def\lb{\lbrace}
\def\lt{\left}
\def\rt{\right}
%%%%%%%%%%%%%%%%%%%%%%sinh
\def\sh{\sinh}
%%%%%%%%%%%%%%%%%%%%%%%%%
%%%%%%\ldots low %%%\vdots vertical
%%%%%%\cdots center%%%%%ddots diagonal

\def\tri{\triangle}
\def\ot{\otimes}
\def\bot{\bigotimes}
\def\op{\oplus}
\def\bop{\bigoplus}

\def\rtarr{\rightarrow}
\def\Rtarr{\Rightarrow}
\def\Ltarr{\Leftarrow}
\def\ltarr{\leftarrow}
\def\lrtarr{\leftrightarrow}
\def\mpto{\mapsto}

\def\dg{\dagger}
\def\apx{\approx}
\def\eqv{\equiv}
\def\sbs{\sebset}
\def\sbsq{\sebseteq}
\def\smq{\simeq}

\def\psd{\psi ^{\dagger}}
\def\Psd{\Psi ^{\dagger}}
\def\psb{\bar{\psi}}
\def\psbd{\bar{\psi} ^{\dagger}}
\def\psp{\psi ^{\prime}}
\def\pspd{{\psi ^{\prime}}^{\dagger}}

%%%%%%%%%%%%%%%%%%%%%%%%%%%%%%%%%%%%%%%%%%%%%%%%%%%%%%%%%%%%%%%%%
%    End of Yao's personal definition                             %
%%%%%%%%%%%%%%%%%%%%%%%%%%%%%%%%%%%%%%%%%%%%%%%%%%%%%%%%%%%%%%%%%

\title{\Large $\mathbb Z_N$ symmetric chiral Rabi model: a new $N$-level system}

\author{Yao-Zhong Zhang}
\email{yzz@maths.uq.edu.au}
\affiliation{School of Mathematics and Physics,
The University of Queensland, Brisbane, Qld 4072, Australia}

\begin{abstract}
We present a new tractable quantum Rabi model for $N$-level atoms by extending the $\mathbb Z_2$ symmetry
of the two-state Rabi model. The Hamiltonian is $\mathbb Z_N$ symmetric and allows the parameters in
the level separation terms to be complex while remaining hermitian. This latter property means that the 
new model is {\em chiral}, which makes it differ from any existing $N$-state Rabi models in
the literature. The $\mathbb Z_N$ symmetry provides partial diagonalization
of the general Hamiltonian. The exact isolated (i.e. exceptional) energies of the model have the
Rabi-like form but are $N$-fold degenerate. For the three-state case ($N=3$), we obtain three
transcendental functions whose zeros give the regular (i.e. non-exceptional) energies of the model.

\end{abstract}


%02.30.lk Integrable systems

%03.65.Fd - Algebraic methods
%03.65.Ge Solutions of wave equations: bound state
%03.75.-b Matter waves (for atoms)
%03.75.Kk Dynamical properties of condensates collective and hydrodynamic excitations, superfluid flow
%03.75.Mn Multicompoment condensates, spinor condensates

%05.30.Jp Boson systems

%37.10.Jk Atoms in optical lattice
%37.10.Gh Atoms traps and guides

%42.50.-p - Quantum optics
%42.50.Pq - Cavity quantum electrodynamics; micromasers 

%67.85.-d Ultracold gases, trapped gases
%67.85.De Dynamical properties of condensates, excitations, superfluid flow
%67.85.Fg Multicomponent condensates, spinor condesates
%67.85.Hj Bose-Einstein condensation in optical lattice

\maketitle

\section{Introduction}\label{introduction}
Matter-light interactions are ubiquitous in nature. In modern physics, they are modeled by 
systems of atoms interacting with boson modes (i.e. spins coupled to harmonic oscillators). 
One of the best-known spin-boson systems is the phenomenological quantum Rabi model 
\cite{Fulton61,Schweber67,Wagner84} which continues to be a subject of significant interest 
\cite{Braak11}-\cite{Moroz14}. This model describes 
the interaction of a two-level atom with a cavity mode of quantized electromagnetic field, i.e.
a single spin-1/2 particle coupled to a harmonic oscillator. Due to the simplicity of its Hamiltonian, the
Rabi model has served as the theoretical basis for understanding the interactions between matter and
light, and has found a variety of applications ranging from quantum optics \cite{Vedral06}, 
solid state semiconductor systems \cite{Khitrova06}, molecular physics \cite{Thanopulos04}
to quantum information \cite{Romero12}. With experimental techniques now available to access ultra-strong
atom-cavity coupling regimes \cite{You11}, there is also much ongoing interest in experimental 
realizations of the Rabi interactions in both circuit and cavity quantum electrodynamics (QED)
\cite{Englund07}-\cite{Chang12}. 

The quantum Rabi model is the simplest spin-boson system without the rotating wave approximation (RWA).
To understand more sophisticated spin-boson interactions, e.g. multi-state atom-cavity interaction
\cite{Scully06,Bianchetti10,Srinivasan11},  there is a need for extensions of
the two-level Rabi model.  In this regard, let us mention three research directions which have attracted
significant attention. One is the study of the Dicke model which couples $N$ two-level systems 
to a single radiation mode and is relevant to experimental realization and applications in quantum 
computing. There is a vast literature on the Dicke model. Analytic solutions for the $N=2, 3$ Dicke models
have recently been studied in \cite{Chilingaryan13,Peng13,Braak13}.
Another one concerns models of a two-level atom interacting with multi harmonic 
modes or with a higher-order harmonic generation.
Examples include the two-mode \cite{Zhang13a} and 2-photon \cite{Emary02} Rabi models 
recently solved analytically in \cite{Zhang13c,Moroz-Zhang14}.
These models can be experimentally realized  in circuit QED systems \cite{Niemczyk10} and have
established applications in e.g. Rubidum atoms \cite{Brune87} and quantum dots \cite{Valle10,Ota11}.
The third direction is to consider multi-state atom-cavity interactions, i.e.  multi-level atoms coupled to
harmonic boson modes. Most previous multi-level extensions \cite{Kus86,Wagner89,Eiermann96,Rapp97}
are neither tractable nor applicable to atom-cavity systems. 
Recently the author in \cite{Albert12} proposed a tractable quantum Rabi model for $N$-state atoms. 
%That model is not $\mathbb Z_N$ symmetric. 

In this Letter we introduce a different, tractable quantum Rabi model for $N$-level atoms.
The Hamiltonian of the new model is $\mathbb Z_N$ symmetric, extending the $\mathbb Z_2$
symmetry of the two-state model. One of the unique features to our model is that 
it allows the parameters in the level-splitting terms ($\alpha_m$ in (\ref{N-state-RabiH1}) below)
to be complex while keeping the Hamiltonian hermitian. It is therefore a {\em chiral} system
\cite{Chiral-model} and is referred to as $\mathbb Z_N$ symmetric $N$-state chiral Rabi model. 
The $\mathbb Z_N$ symmetry provides a partial diagonalization of the Hamiltonian. 
%Similar to the Rabi case,
%the spectrum of the $N$-state model consists of two parts: exceptional and regular spectra.
It is found that the exact isolated (i.e. exceptional) energies of the model have the Rabi-like 
form but are $N$-fold degenerate. They correspond to polynomial solutions of the Schr\"odinger
equation and appear when the model parameters satisfy certain constraints. 
%Majority part of the energy spectrum is regular (i.e. non-exceptional). 
For the three-state case ($N=3$), we analytically determine three transcendental functions whose zeros 
give the regular energies of the system. Our results pave the way for
applications to multi-level atom-cavity experiments.

%\setcounter{section}{1}
%\setcounter{equation}{0}

%%%%%%%%%%%%%%%%%%%%%%%%%%%%%%%%%%%%%%%%%%%%%%%%%%%%%%%%%%%%%%%%%%%%%%%%%%%%%%%%%%%%%%%%%%%%%%%%%%%%%%%%%%%%%%%%%%%%%%%%%%%%%%%
\section{$\mathbb Z_N$ symmetric Rabi Hamiltonian}
The two-state Rabi Hamiltonian is $\mathbb Z_2$ symmetric. So the most natural $N$-state generalization
of the Rabi model would be given by a Hamiltonian with $\mathbb Z_N$ symmetry. We can proceed in the
following intuitive and mathematically rigorous way. Similar to the Rabi
case where a two-level atom is modeled by spins with two states (Pauli matrices $\sigma_z$ and $\sigma_x$),
we model  an $N$-level atom by ``spins" with $N$ states.
Then the Hilbert space of the $N$-state atom is the $N$-dimensional vector space ${\mathbb  C}^N$.
Let $Z$ and $X$ be the basic operators which generalize the
Paul matrices $\sigma_z$ and $\sigma_x$ to ${\mathbb  C}^N$, respectively. 
Instead of anti-commutation relations, these operators satisfy \cite{Sylvester1909}
\beqa
&&Z^N=X^N=1,~~~~Z^\dagger=Z^{N-1},~~~~X^\dagger=X^{N-1},\n
&&ZX=\omega\,XZ,~~~~~\omega=e^{2\pi i/N}.\label{ZX-relations}
\eeqa
It is useful to keep in mind some explicit representations of these operators. Diagonalizing the 
the operator $Z$ gives $Z={\rm diag}(1, \omega,\omega^2,\cdots, \omega^{N-1})$ and
$X_{l,m}=\delta_{l,m+1\,(mod ~ N)}$. On the other hand, from (\ref{ZX-relations}) we have 
$X^\dagger Z=\omega Z X^\dagger$. Thus the representation in which $X$ is diagonal is given by
$X^\dagger= {\rm diag}(1, \omega,\omega^2,\cdots, \omega^{N-1})$ and $Z_{l,m}=\delta_{l,m+1\,(mod ~ N)}$.
This latter representation is useful in what follows.
%\beq
%Z=\left(
%\begin{array}{ccccc}
%1 & 0 & 0 & \cdots & 0\\
%0 & \omega & 0 & \cdots & 0\\
%0 & 0 & \omega^2 & \cdots & 0\\
%  &   &   & \ddots & \\
%0 & 0 & 0 & \cdots & \omega^{N-1}\\
%\end{array}
%\right ),~~~
%X=\left(
%\begin{array}{ccccc}
%0 & 0  & \cdots & 0 & 1\\
%1 & 0  & \cdots & 0 & 0\\
%0 & 1  & \cdots & 0 & 0\\
%  &    & \ddots &   & \\
%0 & 0 & \cdots  & 1 & 0\\
%\end{array}
%\right ).
%\eeq

Then the most natural and mathematically manageable $N$-state generalization of the two-state Rabi model
can be obtained by replacing the Pauli matrices in the latter model with the ``spins" with $N$
states. It is convenient to label the $N$-states by $1, \,\omega,\,\omega^2,\,\cdots,\,\omega^{N-1}$. 
We thus arrive at the following Hamiltonian of a $N$-level atom interacting with a boson mode, 
\beq
H_N=\Omega b^\dagger b+\Delta\sum_{m=1}^{N-1}\alpha_m\,Z^m+\lambda \left(X^\dagger b^\dagger+X b\right),
  \label{N-state-RabiH1}
\eeq
where $\Omega, \Delta, \lambda$ are real parameters, and the couplings $\alpha_m$ are
complex and obey $\alpha_m^\dagger =\alpha_{N-m}$ in order for the Hamiltonian to be hermitian. 
The above $H_N$ is the simplest possible, hermitian and $N$-state generalization of the 
two-state Rabi Hamiltonian. For $N=2$, the Hamiltonian (\ref{N-state-RabiH1}) simplifies to the 
well-studied two-state Rabi model,
\beq
H_2=\Omega b^\dagger b+\Delta \sigma_z+\lambda \sigma_x(b^\dagger+b).\label{RabiH}
\eeq

Let us state clearly that our model differs from the one proposed in
\cite{Albert12} in several important aspects. Firstly, the atom-cavity interaction term of the model 
in \cite{Albert12} (see eq.(6) of that paper) is, in our notation, $\lambda( X^\dagger b+X b^\dagger)$,
while it is $\lambda( X^\dagger b^\dagger+X b)$ in our model. Although both of them reduce to the same 
Rabi model interaction term, for $N\geq 3$ they represent quite different atom-cavity interactions 
since $X$ is non-hermitian. It is interesting that the Rabi model interaction has two different
$N$-state generations.  Secondly, perhaps more importantly, the parameters $\alpha_m$ in our model 
(\ref{N-state-RabiH1}) are in general complex (while the level-splitting parameters in \cite{Albert12} are real).  
%Traditionally the parameter $\alpha_m$ in (\ref{N-state-RabiH1}) are chosen to be real. 
%However, 
Allowing these parameters to be complex may result in interesting behaviour not possible in the
Rabi model. To understand this, consider the three-state system $N=3$ and let
$\alpha_1=\alpha_2^\dagger=e^{i\phi}$, where $\phi$ is a real parameter. Then we obtain
\beq
H_3=\Omega b^\dagger b+\Delta\left(e^{i\phi} Z+e^{-i\phi} Z^\dagger\right)
   +\lambda(X^\dagger b^\dagger+X b).\label{3-state-RabiH1}
\eeq
There are thus three physical important parameters in $H_3$: $\Omega/\lambda, \Delta/\lambda$ and $\phi$.
When the phase $\phi=0$ so that $\alpha_1$ and $\alpha_2$ are real, the Hamiltonian is invariant if
$Z$ and $Z^\dagger$ are interchanged. For $\phi\neq 0$, the Hamiltonian is no-longer invariant if
$Z$ is interchanged with $Z^\dagger$. This means that spatial parity symmetry in any direction is
broken. For this reason, our model with non-zero $\phi$ is  {\em chiral} \cite{Chiral-model}. 

Working in the representation in which $X$ is diagonal, 
$X^\dagger={\rm diag}(1, \omega, \omega^2), Z=\left(\begin{array}{ccc}
0&0&1\\
1&0&0\\
0&1&0\\
\end{array}\right)$, where $\omega=e^{i 2\pi/3}$, 
we can express (\ref{3-state-RabiH1}) in the explicit matrix form
\begin{widetext}
\beq
H_3=\left(
\begin{array}{ccc}
\Omega b^\dagger b+\lambda(b^\dagger+b) & \Delta e^{-i\phi} & \Delta e^{i\phi}\\
\Delta e^{i\phi} & \Omega b^\dagger b+\lambda(\omega b^\dagger+\omega^2 b) & \Delta e^{-i\phi}\\
\Delta e^{-i\phi} & \Delta e^{i\phi} & \Omega b^\dagger b+\lambda(\omega^2 b^\dagger+\omega b)\\
\end{array}
\right).\label{3-state-RabiH2}
\eeq
\end{widetext}

The Hamiltonian (\ref{N-state-RabiH1}) is invariant under the cyclic group $\mathbb Z_N$ and the corresponding
symmetry generator is
\beq
\Pi=Z\,e^{i\frac{2\pi}{N}b^\dagger b},
\eeq
which is an operator acting in the Hilbert space ${\mathbb C}^N\otimes {\cal H}_b$. Throughout
${\cal H}_b$ denotes the Hilbert space of the boson degree of freedoms. 
Indeed it can be shown that $\Pi$ satisfies $\Pi^N=1$ and commutes with the Hamiltonian, $[H_N, \Pi]=0$.
Thus ${\mathbb  C}^N\otimes {\cal H}_b$ splits into $N$
invariant subspaces $|\omega^{k-1}\ra\otimes {\cal H}_b,~ k=1,2, \cdots, N$ labeled by the eigenvalues
 $1, \omega, \omega^2,\cdots, \omega^{N-1}$ of $Z$. 
This invariance can be used to partially diagonalize $H_N$ via the generalized Fulton-Gouterman 
transformation $U_N$ \cite{Wagner84},
\beq
U_N=\frac{1}{\sqrt{N}}\sum_{\gamma=1}^N\sum_{r=1}^N\,\omega^{(r-1)(\gamma-1)}\,|r\ra \la\gamma |\,R_r,
   \label{FG-transformation1}
\eeq
where $R_r=R^{r-1}$ with $R=e^{i\frac{2\pi}{N}b^\dagger b}$. For example, when $N=3$, we have from
(\ref{FG-transformation1})
\beq
U_3=\frac{1}{\sqrt{3}}\left(
\begin{array}{ccc}
1 & 1 & 1\\
R & \omega R & \omega^2 R\\
R^2 & \omega^2 R^2 & \omega R^2\\
\end{array}
\right) , ~~~~R=e^{i\frac{2\pi}{3}b^\dagger b}.\label{FG-tranformation-for-H3}
\eeq
Then it can be checked that
\beq
U^\dagger_3 H_3 U_3=\left(
\begin{array}{ccc}
H^{(0)} & 0 & 0 \\
0 & H^{(1)} & 0 \\
0 & 0 & H^{(2)}\\
\end{array}
\right),
\eeq
where 
\beq
H^{(s)}=\Omega b^\dagger b+\lambda(b^\dagger+b)+\Delta e^{-i\phi} \omega^sR
    + \Delta e^{i\phi} \omega^{2s}R^2
\eeq
%\beqa
%&&H_1=\Omega b^\dagger b+\lambda(b^\dagger+b)+\Delta e^{i\phi} R+ \Delta e^{-i\phi} R^2,\n
%&&H_\omega=\Omega b^\dagger b+\lambda(b^\dagger+b)+\omega \Delta e^{i\phi} R
%      +\omega^2\Delta e^{-i\phi} R^2,\n
%&&H_{\omega^2}=\Omega b^\dagger b+\lambda(b^\dagger+b)+\omega^2 \Delta e^{i\phi} R
%    +\omega\Delta e^{-i\phi} R^2\n
%\eeqa
act in three mutually orthogonal subspaces ${\cal H}_b\otimes |\omega^s\ra$  with fixed eigenvalue
$\omega^s$ of $Z$. Here and throughout $s=0,1, 2$.

%%%%%%%%%%%%%%%%%%%%%%%%%%%%%%%%%%%%%%%%%%%%%%%%%%%%%%%%%%%%%%%%%%%%%%%%%%%%%%%%%%%%%%%%%%%%%%%%%%%%%%%%%%%%%%%%%%%%%%%%

\section{Exact isolated energies}
We represent the continuous boson degree of freedom $b^\dagger, b$ as differential operators in the
Bargmann-Hilbert space ${\cal B}$ of entire functions which is isomorphic to 
${\cal H}_b\otimes {\mathbb C}^N$. The scalar product of any two elements $f(z), g(z)$ in the 
Bargmann-Hilbert space is given by 
$
%\beq
(f,g)=\int\,\overline{f(z)}\,g(z)\,d\mu(z),
%\eeq
$
where $d\mu(z)=\frac{1}{\pi}e^{-|z|^2}\,dx\,dy$. 
An orthonormal basis of the Bargmann-Hilbert space is provided by the monomials $\{z^n/\sqrt{n!}\}$.
So if $f(z)=\sum_{n=0}^\infty c_n z^n$, then
$||f||^2=\sum_{n=0}^\infty\,|c_n|^2\,n!$ and $f(z)$ is entire iff this sum converges. 
With respect to the orthonormal basis, we have the Bargmann realization $b^\dagger = z$ and 
$b=\frac{d}{dz}$. In this section, $\omega=e^{i 2\pi/N}$.

In the Bargmann representation the eigenvectors of (\ref{N-state-RabiH1}) can be written as $N$-component,
$z$-dependent vectors $\psi(z)=(\psi_1(z), \cdots,\psi_N(z))^t$, 
satisfying the time-independent matrix Schr\"odinger equation
\beq
\left[(\Omega z+\lambda X)\frac{d}{dz}+\lambda z X^\dagger+\Delta\sum_{m=1}^{N-1}\alpha_m\,Z^m\right]\psi(z)
   =E\psi(z).\label{N-state-Rabi-SchEq1}
\eeq
Working in the representation in which $X$ is diagonal, we write this equation in terms of the components
$\psi_k(z)$,
%\begin{widetext}
\beqa
\left(\Omega z+\lambda (\omega^*)^{k-1}\right)\frac{d\psi_k}{dz}
  &=&\left(E-\lambda \omega^{k-1}\,z\right)\psi_k\n
& &-\Delta\sum_{m=1}^{N-1}\alpha_m   \sum_{l=1}^{N}\left(Z^m\right)_{kl}\psi_l,\n
    \label{N-state-Rabi-SchEq2}
\eeqa
%\end{widetext}
where $k=1,2,\cdots,N$.
This is a system of $N$ coupled differential equations of Fuchsian type.  In general it has $N$
independent solutions. The required solution must be analytic in the whole complex plane, i.e. is entire,
in order for $E$ to belong to the spectrum of the system.   The singular points of the
system are at $z=-\frac{\lambda}{\Omega}(\omega^*)^{k-1},~k=1,2,\cdots,N$. 
Assuming that $\psi(z)$ behaves like
$\left(z+\frac{\lambda}{\Omega}(\omega^*)^{k-1}\right)^\rho$ in the vicinity of each
$z=-\frac{\lambda}{\Omega}(\omega^*)^{k-1}$, we are lead to the following indicial equation:
\beq
\rho-\frac{E}{\Omega}-\frac{\lambda^2}{\Omega^2}=0
\eeq
for all $k=1,2,\cdots,N$, where we have used the fact that $\left(Z^m\right)_{kk}=0$ for $m=1,2,\cdots,N-1$.
So for all $N$ independent solutions $\psi_k(z)$ to be analytic at the singular points 
$z=-\frac{\lambda}{\Omega}(\omega^*)^{k-1}$, we must have 
\beq
E=\Omega\left({\cal N}-\frac{\lambda^2}{\Omega^2}\right),~~~~~{\cal N}=0,1,2,\cdots.
\eeq
This gives the exact isolated energies of the $N$-state Rabi model. These energies
have the Rabi-like form but are $N$-fold degenerate. They correspond to polynomial solutions 
(wavefunctions) of the Schr\"odinger equation (\ref{N-state-Rabi-SchEq2}) and
appear when the model parameters fulfill certain constraints.

%%%%%%%%%%%%%%%%%%%%%%%%%%%%%%%%%%%%%%%%%%%%%%%%%%%%%%%%%%%%%%%%%%%%%%%%%%%%%%%%%%%%%%%%%%%%%%%%%%%%%%%%%%%%%%%%%%%%%%%%%%%%%%%

\section{Regular energies}
Here for the purpose of illustration, we will consider the three-state ($N=3$) case.
Generalization to the $N$-state model is straightforward. Throughout this section
$\omega=e^{i 2\pi/3}$.

In the Bargmann representation, $R$ in (\ref{FG-tranformation-for-H3}) can be realized as
$R=e^{i\frac{2\pi}{3}z\frac{d}{dz}}$, which acts on elements $f(z)$ of ${\cal B}$ as
$(R\,f)(z)=f(\omega z)$. Thus we can express the Hamiltonians $H^{(s)},~s=0,1,2,$ as 
the differential operators in ${\cal B}$
\beq
H^{(s)}=(\Omega z+\lambda)\frac{d}{dz}+\lambda z+\Delta e^{-i\phi} \omega^{s+z\frac{d}{dz}}
    + \Delta e^{i\phi} \omega^{2s+2z\frac{d}{dz}}.
\eeq
%\begin{widetext}
%\beqa
%&&H_1=(\Omega z+\lambda)\frac{d}{dz}+\lambda z+\Delta e^{i\phi} \omega^{z\frac{d}{dz}}
%    + \Delta e^{-i\phi} \omega^{2z\frac{d}{dz}},\n
%&&H_\omega=(\Omega z+\lambda)\frac{d}{dz}+\lambda z+\Delta e^{i\phi} \omega^{1+z\frac{d}{dz}}
%    + \Delta e^{-i\phi} \omega^{2+2z\frac{d}{dz}},\n
%&&H_{\omega^2}=(\Omega z+\lambda)\frac{d}{dz}+\lambda z+\Delta e^{i\phi} \omega^{2+z\frac{d}{dz}}
%    + \Delta e^{-i\phi} \omega^{1+2z\frac{d}{dz}}.
%\eeqa
%\end{widetext}
The corresponding time-independent Schr\"odinger equations are
%\begin{widetext}
\beqa
&&\left[(\Omega z+\lambda)\frac{d}{dz}+\Delta e^{-i\phi}\omega^{s+z\frac{d}{dz}}\right.\n
&&~~~~~ \left. + \Delta e^{i\phi} \omega^{2s+2z\frac{d}{dz}}+\lambda z-E^{(s)}\right]\psi^{(s)}(z)=0
    \label{schroedinger-eqn}
\eeqa
%\end{widetext}
for $s=0,1,2$.
%\beqa
%&&\left\{(\Omega z+\lambda)\frac{d}{dz}+\Delta e^{i\phi} \omega^{z\frac{d}{dz}}
%    + \Delta e^{-i\phi} \omega^{2z\frac{d}{dz}}+\lambda z-E_1\right\}\psi_1(z)=0,\n
%&&\left\{(\Omega z+\lambda)\frac{d}{dz}+\Delta e^{i\phi} \omega^{1+z\frac{d}{dz}}
%    + \Delta e^{-i\phi} \omega^{2+2z\frac{d}{dz}}+\lambda z-E_\omega\right\}\psi_\omega(z)=0,\n
%&&\left\{(\Omega z+\lambda)\frac{d}{dz}+\Delta e^{i\phi} \omega^{2+z\frac{d}{dz}}
%    + \Delta e^{-i\phi} \omega^{1+2z\frac{d}{dz}}+\lambda z-E_{\omega^2}\right\}\psi_{\omega^2}(z)=0.
%\eeqa
%\end{widetext}
Here we have written $E^{(s)}$ %$E^{(1)}, E^{(\omega)}, E^{(\omega^2)}$ 
since in general the spectra of $H^{(s)}$ %$H^{(1)}, H^{(\omega)}, H^{(\omega^2)}$ 
are not the same. Solutions to these differential equations must be analytic in the
whole complex plane if $E^{(0)}, E^{(1)}, E^{(2)}$ belong to the spectra of $H^{(0)},
 H^{(1)}, H^{(2)}$, respectively. In other words, we are seeking solutions of the form
\beq
\psi^{(s)}(z)=\sum_{n=0}^\infty\, K_n^{(s)}(E^{(s)})\,z^n,\label{series-solution}
\eeq
which converge in the whole complex plane, i.e. are entire.

Substituting (\ref{series-solution}) into (\ref{schroedinger-eqn}), we obtain the 3-term recurrence
relations
\beqa
&& K^{(s)}_1+A^{(s)}_0\,K^{(s)}_0=0,\n  
&& K^{(s)}_{n+1}+A^{(s)}_n\,K^{(s)}_n+B^{(s)}_n\,K^{(s)}_{n-1}=0,~~~n\geq 1,\label{3-step1}
\eeqa
where
\beqa
A^{(s)}_n&=&\frac{n\Omega+\Delta e^{-i\phi}\omega^{n+s}
    +\Delta e^{i\phi}\omega^{2(n+s)}-E^{(s)}}{\lambda(n+1)},\n
B^{(s)}_n&=&\frac{1}{n+1}.
\eeqa
The coefficients $A^{(s)}_n, B^{(s)}_n$ have the asymptotic behavior when $n\rightarrow\infty$,
\beq
A^{(s)}_n\sim \frac{\Omega}{\lambda},~~~~~~B_n\sim n^{-1}.
\eeq
Applying the Perron-Kreuser theorem
(i.e. Theorem 2.3 of \cite{Gautschi67}), it follows that for each $s=0,1,2$, the truly 3-term part 
(i.e. the $n\geq 1$ part) of the recurrence relations (\ref{3-step1}) has two linearly 
independent solutions $K^{(s)}_{n,1},
K^{(s)}_{n,2}$ for which, when $n\rightarrow\infty$
\beq
\frac{K^{(s)}_{n+1,1}}{K^{(s)}_{n,1}}\sim -\frac{\Omega}{\lambda},~~~~~~
\frac{K^{(s)}_{n+1,2}}{K^{(s)}_{n,2}}\sim   -\frac{\lambda}{\Omega}\,n^{-1}
\eeq
So $K^{(s)min}_n\equiv K^{(s)}_{n,2}$ for each $s$ value 
is a minimal solution of the truly 3-term part of (\ref{3-step1}). 
The corresponding infinite power series solutions, generated by substituting $K^{(s)min}_n$ 
for the $K^{(s)}_n$'s in (\ref{series-solution}), converge in the whole complex plane, 
i.e. they are entire.

By the Pincherle theorem, i.e. Theorem 1.1 of \cite{Gautschi67}, the ratios of successive elements of
the minimal solution sequences $K^{(s)min}_n,~ s=0,1,2,$ are expressible 
in terms of infinite continued fractions. Proceeding in the direction of increasing $n$, we find
\beq
S^{(s)}_{n}=\frac{K_{n+1}^{(s)min}}{K_n^{(s)min}}=-\frac{B^{(s)}_{n+1}}{~A^{(s)}_{n+1}-}\,
   \frac{B^{(s)}_{n+2}}{~A^{(s)}_{n+2}-}\,
   \frac{B^{(s)}_{n+3}}{~A^{(s)}_{n+3}-}\,\cdots,   \label{continued-fraction}
\eeq
which for $n=0$ gives 
\beq
S^{(s)}_{0}=\frac{K_{1}^{(s)min}}{K_0^{(s)min}}=-\frac{B^{(s)}_{1}}{~A^{(s)}_{1}-}\,
   \frac{B^{(s)}_{2}}{~A^{(s)}_{2}-}\,
   \frac{B^{(s)}_{3}}{~A^{(s)}_{3}-}\,\cdots.   \label{continued-fraction1}
\eeq
Note that the ratio $S^{(s)}_0=\frac{K_1^{(s)min}}{K_0^{(s)min}}$ involve $K_0^{(s)min}$, 
although the above continued fraction expressions are obtained from the truly 3-term part 
of (\ref{3-step1}), i.e the recurrence (\ref{3-step1}) for $n\geq 1$. 
However, for single-ended sequences such as those appearing in the infinite power series
expansions (\ref{series-solution}), the ratios $S^{(s)}_0=\frac{K_1^{(s)min}}{K_0^{(s)min}}$ 
of the first two terms of 
minimal solutions are unambiguously fixed by the $n=0$ part (i.e. the first equation) of 
the recurrence (\ref{3-step1}), namely,
\beq
 S^{(s)}_0=-A^{(s)}_0=-\frac{1}{\lambda}\left[\Delta\left(\omega^s e^{-i\phi}+\omega^{2s}e^{i\phi}\right)
   -E^{(s)}\right].   \label{continued-fraction2}
\eeq
 In general, the $S^{(s)}_0$ computed from (\ref{continued-fraction1}) are not the same as those from 
(\ref{continued-fraction2}) (i.e. (\ref{continued-fraction1}) and  (\ref{continued-fraction2})
are not both satisfied)  for arbitrary values of recurrence coefficients $A^{(s)}_n$ and $B^{(s)}_n$. 
As a result, general solutions to the recurrence (\ref{3-step1}) are dominant and are usually
generated by simple forward recursion from a given value of $K^{(s)}_0$. Physical meaningful
solutions are those that are entire in the Bargmann-Hilbert spaces. They can be obtained if $E^{(s)}$
can be adjusted so that equations (\ref{continued-fraction1}) and (\ref{continued-fraction2})
are both satisfied for each $s$. Then the resulting solution sequences $K^{(s)}_n(E^{(s)})$ 
will be purely minimal and the power series expansions (\ref{series-solution}) will converge 
in the whole complex plane.

Therefore, if we define the functions $F^{(s)}(E^{(s)})=S^{(s)}_0+A^{(s)}_0$ with $S^{(s)}_0$  
given by the continued fraction in (\ref{continued-fraction1}),  then the
 zeros of $F^{(s)}(E^{(s)})$ correspond to the points in the parameter space where the condition 
(\ref{continued-fraction2}) is satisfied. In other words, $F^{(s)}(E^{(s)})=0$ 
yield the eigenvalue equations,
which may be solved for $E^{(s)}$ by standard nonlinear root-search techniques \cite{Leaver86}. 
Only for the denumerable infinite values of $E^{(s)}$ which are the roots of $F^{(s)}(E^{(s)})=0$, do
we get entire wavefunction solutions of the Schr\"odinger differential equations.

%%%%%%%%%%%%%%%%%%%%%%%%%%%%%%%%%%%%%%%%%%%%%%%%%%%%%%%%%%%%%%%%%%%%%%%%%%%%%%%%%%%%%%%%%%%%%%%%%%%%%%%%%%%%%%%%%%%%%%%%%%%%%%%

\section{Conclusions and discussions}\label{summary}
This work introduces a new $\mathbb Z_N$ symmetric $N$-state extension of the two-state Rabi model.
A unique feature to the model is that it allows parameters $\alpha_m$ to be
complex without violating the hermiticity of the Hamiltonian. This is not possible in the Rabi model 
since $\sigma^\dagger_z=\sigma_z$.  Our model is a $N$-state Rabi model with complex level 
separation terms and is thus a {\em chiral system} \cite{Chiral-model}. 
This is one of the main differences between our model and the existing $N$-state models in the literature.
The $\mathbb Z_N$ symmetry can be used to partially diagonalize the Hamiltonian of the model. 
Analytic solutions
of the model has been investigated in the Bargmann-Hilbert space. It is found that the exact isolated
energies have the Rabi-like form but are $N$-fold degenerate. They correspond to polynomial 
wavefunctions and special model parameters. The regular energies are given by zeros of suitable
transcendental functions, similar to the Rabi case. This is shown for the three-state $N=3$ system.

{}From the 3-term recurrence relations (\ref{3-step1}) it is not difficult to show that the wavefunction
expansion coefficients in (\ref{series-solution}) are related to orthogonal polynomials \cite{Chihara78}. 
Thus it is expected that the regular energies of the $N$-state model can be determined as the 
polynomial zeros by a procedure similar to that in \cite{Moroz13,Moroz14}. 

The new $N$-state Rabi Hamiltonian presented in this work is the simplest possible, hermitian and
$\mathbb Z_N$ symmetric generalization of the two-state Rabi model, using a minimal number of system
parameters. More sophisticated extensions are possible. For example, the Hamiltonian of    
another $N$-state generalization which preserves the $\mathbb Z_n$ symmetry has the form
%\begin{widetext}
\beqa
\tilde{H}_N&=&\Omega b^\dagger b+\Delta\sum_{m=1}^{N-1}\alpha_m\,Z^m\n
& &+\lambda\sum_{m=1}^{N-1}\beta_m
  \left[X^m(b^\dagger)^{N-m}+(X^\dagger)^m b^{N-m}\right]\label{another-H}
\eeqa
%\end{widetext}
where $\Omega, \Delta, \lambda$ are real and the couplings $\alpha_m, \beta_m$ are complex and satisfy 
$\alpha^\dagger_m=\alpha_{N-m}$ and $\beta^\dagger_m=\beta_{N-m}$ in order for the Hamiltonian to be
hermitian. It can be checked that this Hamiltonian commutes with the $\mathbb Z_n$ operator $\Pi$.
For $N=2$, (\ref{another-H}) also simplifies to the two-state Rabi model Hamiltonian. 
However, for $N\geq 3$ the Hamiltonian (\ref{another-H})
contains non-linear terms of spin and boson operators.  A detailed analysis of this extension 
is interesting but beyond the scope of this paper.

%\vskip.2in
\begin{acknowledgments}
We would like to thank Victor Albert for critical comments and email conversations, and
Daniel Braak for comments and useful suggestions. We also thank Jacques H.H. Perk for pointing
out a misprint and some references on the chiral Potts model.
This work was partially supported by the Australian Research Council 
through Discovery-Projects grants DP110103434 and DP140101492.
\end{acknowledgments}

\bebb{99}

\bbit{Fulton61}
R.L. Fulton and M. Gouterman, J. Chem. Phys. {\bf 35}, 1059 (1961).

\bbit{Schweber67}
S. Schweber, Ann. Phys. {\bf 41}, 205 (1967).

\bbit{Wagner84}
M. Wagner, J. Phys. A {\bf 17}, 2319 (1984).

\bbit{Braak11}
D. Braak, Phys. Rev. Lett. {\bf 107}, 100401 (2011).

\bbit{Solano11}
E. Solano, Physics {\bf 4}, 68 (2011).

\bbit{Moroz12}
A. Moroz, Europhys. Lett. {\bf 100}, 60010 (2012).
%Comment on ``Integrability of the Rabi model", arXiv:1205.3139v2 [quant-ph].

\bbit{Chen12}
Q.H. Chen, C. Wang, S. He, T. Liu and K.L. Wang, Phys. Rev. A {\bf 86}, 023822 (2012).

\bbit{Zhong13}
H. Zhong, Q. Xie, M.T. Batchelor and C. Lee, J. Phys. A {\bf 46}, 415302 (2013).

\bbit{Tomka13}
M. Tomka, O.E' Araby, M. Pletyukhov and V. Gritsev, arXiv:1307.7876v1 [quant-ph]
 
\bbit{Moroz13}
A. Moroz, Ann. Phys. {\bf 338}, 319 (2013).  

\bbit{Zhang13c}
Y.-Z. Zhang, arXiv:1304.7827v2 [quant-ph].

\bbit{Moroz14}
A. Moroz, Ann. Phys. {\bf 340}, 252 (2014).

\bbit{Vedral06}
V. Vedral, Modern foundations of quantum optics, Imperial College Press, London, 2006.

\bbit{Khitrova06}
G. Khitrova,  H.M. Gibbs, M. Kira, S.W. Koch and A. Scherer, 
Nature Phys. {\bf 2}, 81 (2006).

\bbit{Thanopulos04}
I. Thanopulos, E. Paspalakis and Z. Kis, Chem Phys. Lett. {\bf 390}, 228 (2004).

\bbit{Romero12}
G. Romero, D. Ballester, Y.M. Wang, V. Scarani and E. Solano, Phys. Rev. Lett. {\bf 108}, 120501 (2012).

\bbit{You11}
J.Q. You and F. Nori, Nature {\bf 474}, 589 (2011).

\bbit{Englund07}
D. Englund, A. Faraon, I. Fushman, N. Stoltz, P. Petroff and J. Vuckovi\'c, Nature {\bf 450}, 857 (2007).

\bbit{Niemczyk10}
T. Niemczyk, F. Deppe, H. Huebl, E.P. Menzel, F. Hocke, M.J. Schwarz, J.J. Garcia-Ripoll, D. Zueco, 
T. H\"ummer, E. Solano, A. Marx and R. Gross, Nature Phys. {\bf 6}, 772 (2010). 

\bbit{Casanova10}
J. Casanova, G. Romero, I. Lizuain, J.J. Garcia-Ripoll and E. Solano, Phys. Rev. Lett. {\bf 105}, 
263603 (2010).

\bbit{Schuster10}
D. Schuster, A. Fragner, M. Dykman, S. Lyon and R. Schoelkopf, Phys. Rev. Lett. {\bf 105}, 040503 (2010).

\bbit{Longhi11}
S. Longhi, Opt. Lett. {\bf 36}, 3407 (2011).

\bbit{Crespi12}
A. Crespi, S. Longhi and R. Osellame, Phys. Rev. Lett. {\bf 108}, 163601 (2012).

\bbit{Chang12}
D.E. Chang, L. Jiang, A.V. Gorshkov and H.J. Kimble, arXiv:1201.0643v3 [quant-ph].

\bbit{Scully06}
M. Scully, E. Fry, C. Ooi and K. W\'odkiewicz, Phys. Rev. Lett. {\bf 96}, 010501 (2006).

\bbit{Bianchetti10}
R. Bianchetti, S. Filipp, M. Baur, J. Fink, C. Lang, L. Steffen, M. Boissonneault, A. Blais and A. Wallraff,
Phys. Rev. Lett. {\bf 105}, 223601 (2010).

\bbit{Srinivasan11}
S. Srinivasan, A. Hoffman, J. Gambetta and A. Houck, Phys. Rev. Lett. {\bf 106}, 083601 (2011).

\bbit{Chilingaryan13}
S.A. Chilingaryan and B.M. Rodriguez-Lara, J. Phys. A {\bf 46}, 335301 (2013). 

\bbit{Peng13}
J. Peng, Z. Ren, D. Braak, G. Guo, G. Ju, X. Zhang and X. Guo, arXiv:1312.7610 [quant-ph]. 

\bbit{Braak13}
D. Braak, J. Phys. B {\bf 46}, 224007 (2013).

\bbit{Zhang13a}
Y.-Z. Zhang, J. Math. Phys. {\bf 54}, 102104  (2013).

\bbit{Emary02}
C. Emary and R.F. Bishop, J. Phys. A {\bf 35}, 8231 (2002).

\bbit{Moroz-Zhang14}
A. Moroz and Y.-Z. Zhang, in preparation.

\bbit{Brune87}
M. Brune,  J.M. Raimond, P. Goy, L. Davidovich and S. Haroche, 
Phys. Rev. Lett. {\bf 59}, 1899 (1987).

\bbit{Valle10}
E. del Valle,   S. Zippilli, F.P. Laussy, A. Gonzalez-Tudela, G. Morigi and C. Tejedor,
Phys. Rev. B {\bf 81}, 035302 (2012).

\bbit{Ota11}
Y. Ota,  S. Iwamoto, N. Kumagai and Y. Arakawa, %Spontaneous two photon emission from a single quantum dot,
 arXiv:1107.0372v1 [quant-ph].

\bbit{Kus86}
M. Ku\`s and M. Lewenstein, J. Phys. A {\bf 19}, 305 (1986).

\bbit{Klenner87}
N. Klenner and J. Weis, J. Phys. A {\bf 20}, 1155 (1987).

\bbit{Wagner89}
M. Wagner and A. Kongeter, J. Chem. Phys. {\bf 91}, 3036 (1989).

\bbit{Eiermann96}
H. Eiermann and M. Wagner, J. Chem. Phys. {\bf 105}, 6713 (1996).

\bbit{Rapp97}
M. Rapp and M. Wagner, J. Phys. A {\bf 30}, 2811 (1997).

\bbit{Albert12}
V.V. Albert, Phys. Rev. Lett. {\bf 108}, 180401 (2012).

\bbit{Chiral-model}
The notion ``chiral" first appeared in work on the chiral clock/Potts model, 
which is the $Q$-state generalization of the Ising model. There is a big literature on
chiral Potts model. Here we list a few early papers on the topic.
S. Ostlund, Phys. Rev. B {\bf 24}, 398 (1981); D.A. Huse, A.M. Szpilka and M.E. Fisher,
Physica A {\bf 121}, 363 (1983); G. von Gehlen and V. Rittenberg,
Nucl. Phys. B {\bf 257}, 351 (1985);
H. Au-Yang, B.M. McCoy, J.H.H. Perk, S. Tang S and M.L. Yan, Phys. Lett. A {\bf 123}, 219 (1987);
R.J. Baxter, Phys. Lett. A {\bf 140}, 155 (1989).

\bbit{Sylvester1909}
The non-hermitian generalization of Pauli matrices to higher dimensions was obtained by
Sylvester in 1882. See e.g. The Collected Mathematics Papers of James Joseph Sylvester
(Cambridge University Press, 1909). The $Z, X$ notation was first used by Au-Yang et al in
\cite{Chiral-model}.

\bbit{Gautschi67}
W. Gautschi, SIAM Rev. {\bf 9}, 24 (1967).

\bbit{Leaver86}
E.W. Leaver, J. Math. Phys. {\bf 27}, 1238 (1986).

\bbit{Chihara78}
T.S. Chihara, An introduction to orthogonal polynomials, Gordon and Breach, New York, 1978.

\eebb

\end{document}